\documentclass{elsart}
\usepackage{amssymb}
\usepackage{amsmath}\allowdisplaybreaks
\usepackage{graphicx}
\usepackage[normalem]{ulem}
\usepackage[dvips]{color}
\begin{document}
\begin{frontmatter}

\title{Neutron stars with spin polarized self-interacting dark matter}
\author {Zeinab Rezaei\thanksref{info1}}

\address{Department of Physics, Shiraz
University, Shiraz 71454, Iran}
\address{Biruni Observatory, Shiraz
University, Shiraz 71454, Iran}
\thanks[info1]{zrezaei@shirazu.ac.ir}

\begin{abstract}
Dark matter, one of the important portion of the universe,
 could affect the visible matter in neutron stars.
An important physical feature of dark matter
is due to the spin of dark matter particles.
Here, applying the piecewise polytropic equation of state for the neutron star matter and the equation of state of spin polarized self-interacting dark matter, we investigate the structure of
neutron stars which are influenced by the spin polarized self-interacting dark matter.
The behavior of the neutron star matter and dark matter portions for the stars with different
values of the interaction between dark matter particles and spin polarization of
dark matter is considered.
In addition, we present the value of the gravitational redshift of these
stars in different cases of spin polarized and self-interacting dark matter.

\vspace{5mm} \noindent {\it PACS:}
97.60.Jd    
95.35.+d    

\end{abstract}

\begin{keyword}
Neutron stars; Dark matter; Spin polarization
\end{keyword}

\end{frontmatter}


\section{Introduction}

Because of the compactness and high density of compact objects, the accretion of dark matter (DM) particles can
take place on compact stars \cite{Bertone,Kouvaris10}.
The compact objects are sensitive probes of DM and they set constraints on the
properties of DM particles and its density \cite{Bertone}. It has been shown that the heating due to the
accretion of WIMPs onto cool white dwarf stars could be detected \cite{Bertone}.
Self-annihilating neutralino WIMP DM accreted onto neutron stars results in
a mechanism to seed compact objects with long-lived lumps of strange quark matter,
indicating a possible conversion of most of the star into a strange star \cite{Perez-Garcia10}.
This self-annihilation can affect their kinematical properties such as velocity
kicks and rotation patterns \cite{Perez-Garcia12}.
Since even neutron stars located at regions of
low DM density can accrete WIMPs which lead to collapse and form a mini black hole \cite{Goldman,Kouvaris11},
the constraints on the WIMP self-interactions which are stricter than the ones from
the bullet cluster have been derived \cite{Kouvaris12}.
The accretion of millicharged DM onto a neutron
star can result in the expulsion of extra electric charge from the poles of the star
and impede further the rotation of the star yielding braking indices consistent with
the observational results \cite{Kouvaris14}.
A limit on the neutron star survival rate against
transitions to more compact objects has been suggested according to the amount of
decaying DM accumulated in the central regions in neutron stars \cite{Perez-Garcia15}.
This limit also sets constraints on the DM particle decay time.
The collapse of
neutron stars due to capture and sedimentation of DM within their cores
presents a solution for the problems of non-detection of pulsars within the galaxy inner region $\thicksim10\ pc$
and the sources of fast radio bursts \cite{Fuller}. Considering the number of self-interacting but not
self-annihilating DM particles that a neutron star accumulates over its lifetime,
it has been confirmed that the DM self-interactions have a significant role in the rapid accumulation
of DM in the core of neutron star \cite{Guver}.

The compact stars made of fermionic DM have been studied applying the free and
interacting Fermi gas model for the DM \cite{Narain}. The results show that a
unique mass-radius relation for compact stars made of free fermions exists which is
independent of the fermion mass. Besides, the mass-radius relation for compact stars
with strongly interacting fermions indicates that the radius remains constant for a wide
range of compact star masses.
 The repulsion of DM in neutron star leads to form DM halo \cite{Xiang}.
This repulsion results in disappearing the density dependencies of nuclear symmetry energy
which this also leads to difference in particle number density distributions in DM admixed
neutron stars
and consequently in star radii \cite{Xiang}.
Since the
DM halo of star may extend around the star, the mass of halo can result in
the alternative gravitational effects \cite{Li,Rezaei}.
The stellar structure \cite{Sandin,Ciarcelluti,Xiang},
temperature \cite{Kouvaris8}, linear and angular momentum \cite{Perez-Garcia12} of neutron star are affected by DM.

The spin state of DM particles and the properties of DM related to
the spin polarization of these particles have been explored in the last decade.
The spin of the mother particle and the DM particle
in the center of momentum system of a decaying particle can be specified \cite{Christensen}.
Theoretically, it has been confirmed that a spin one half
matter field with mass dimension one and the dominant interaction via Higgs
can be a candidate for DM \cite{Ahluwalia06,Ahluwalia012}.
A DM
candidate particle of spin $3/2$ with neutrino-like Standard Model strength interactions has been
introduced \cite{Savvidy}. This particle can
couple to the nucleon via Z-exchange and it may lead to large spin-independent and spin-dependent
cross sections for a Dirac or Majorana particle, respectively.

A model independent analysis considering the spin-dependent interactions with both protons and
neutrons has been studied \cite{Barnich}. In the detection of DM,
$^{19}F$ is the most useful particle to detect the spin-dependent interactions \cite{Tovey}.
The spin-dependent WIMP interactions on $^{19}F$ have been searched by the PICASSO experiment at Sudbury
Neutrino Observatory LAB using the superheated droplet technique \cite{Archambault}. The cold DM
has been examined by the direct detection of WIMPs based on the spin-dependent
interactions with nuclei \cite{Martel}.
There exists
some mechanisms of interferences that lead to suppression of the spin-independent interaction
in the scattering of scalar DM with nucleus compared to the spin-dependent interaction \cite{Martinez}.
It has been argued that by considering the Maxwellian speed distribution, the spin-dependent interactions
become the dominant source of scattering around the interference regions.
In addition,
for a collision with a known speed, the dominance of the spin interaction presents stringent
limits of the WIMP mass around the interference point. By modeling three stars including DM
energy transport and applying asteroseismic diagnostics, the indications
limiting the effective spin-dependent DM-proton coupling for masses of a few GeV was found \cite{Martins}.
In that work, using the observational data and the results of stellar models including DM
energy transport, it has been tried to constrain the properties of low-mass asymmetric DM with an effective
spin-dependent coupling.
The spin contribution on the cross section of natural DM candidate from supersymmetry
has been investigated \cite{Vergados}. Models which predict a substantial fraction of higgsino lead
to a relatively large spin induced cross section due to the Z-exchange \cite{Vergados}.
The dependency
of DM scattering on the intrinsic spin of DM particles has been studied \cite{Barger}.
The general formulas for spin-dependent cross sections for the scattering of WIMPS with intrinsic
spin $0$, $1/2$, $1$, and $3/2$ have also been considered in Ref. \cite{Barger}. The sensitivity
of the MIMAC-He3 detector for supersymmetric DM search to neutralinos ($M_{\tilde{\chi}}\gtrsim 6\ GeVc^{-2}$) has been
studied via spin-dependent interaction with $^3He$ \cite{Moulin}. This sensitivity leads to complementarity
of MIMAC-He3 with ongoing experiments. The isotope $^{73}Ge$ can provide probe to the spin-dependent
couplings of WIMPs with the neutrons \cite{Lin}. The improved limits on spin-independent and spin-dependent
couplings of low-mass WIMP DM with a germanium detector have been presented \cite{Lin}.
The models for the direct DM detection which are detectable via spin-dependent interactions
have been explored \cite{Freytsis}. The findings verify that most models with detectable spin-dependent
interactions generate detectable spin-independent interactions. Different elastic spin-dependent
operators have been studied in the detection and solar capture of WIMP \cite{Liang}.
It has been concluded that the efficiency of the detection strategies depends on the spin-dependent operators.

The size
of the bound WIMP population for the DM bound to the solar system by solar capture
depends on the WIMP mass $m$, spin-independent cross section,
and spin-dependent cross section \cite{Peter}.
The central temperature of the Sun and the resulting $^8B$ neutrino flux decrease in the
Models of DM with large spin-dependent interactions and an intrinsic asymmetry that
prevents post freeze-out annihilations \cite{Cumberbatch}.
The constraints on the spin-dependent cross section of
asymmetric fermionic DM WIMPs based on the existence of compact stars in globular
clusters have been investigated \cite{Kouvaris11}. It has been confirmed that asymmetric WIMP
candidates with only spin-dependent interactions trapped during the lifetime of the progenitor can
thermalize inside the white dwarf.
Moreover, the characteristics of the DM capture rate by stars
are very different for the spin-dependent and spin-independent DM particle-nucleon scattering cross
sections \cite{Lopes11}.

Some other studies have been considered to explore the spin-dependent and magnetic properties
of DM particles.
Considering a neutral Dirac fermion as a DM candidate, it has been shown that
the elastic scattering is due to a spin-spin interaction \cite{Heo}.
The models of inelastic DM for the DM direct detection
 experiments in which iodine with its large magnetic moment is used have
 been investigated \cite{Chang}. In that study, the dipole
moments for the WIMP have been applied
with the conventional magnetism and also dark magnetism and
both magnetic-magnetic and magnetic-electric scattering.
Moreover, a model of multi-component DM with magnetic moments has been
presented to describe the
$130\ GeV$ gamma-ray line hinted by the Fermi-LAT data \cite{Gu}.
Another aspects of the magnetic properties of DM particles has been suggested in Ref. \cite{Berezhiani}.
In that work, it has been argued that the unconventional properties of DM may create
the galactic magnetic fields.
The effect of supernova explosions on magnetized DM halos has been explored
applying a set of high resolution simulations \cite{Seifried}.
In an interesting recent research, the detection of magnetized quark nuggets
as a candidate for DM has been investigated \cite{VanDevender}.
They have calculated the flux of electromagnetic radiation from electrons
 which is swept up by the magnetic
field of quark-nuggets and their synchrotron radiation from a
magnetized quark nugget interaction with the galactic magnetic field.

On the other hand, it has been argued that
for the ordinary matter, the spin-dependent interactions can
result in the spontaneous spin polarized systems \cite{Kutschera,Klironomos,Sablikov,Dolgopolov}. In nuclear
matter, for non-localized protons, there exists a threshold
value of the spin interaction above which the system can develop
a spontaneous polarization \cite{Kutschera}. Besides, it has
been found that strong interactions between electrons lead to a
ferromagnetic ground state in a certain range of electron  densities \cite{Klironomos}.
The phase transition with spontaneous breaking the spin symmetry due to exchange
interaction in electron systems has been also reported in Ref. \cite{Sablikov}.
In addition, the spin polarization increases with increasing the
interaction parameter \cite{Sablikov}. Moreover, it has been shown that a
spin polarized electron system localizes at electron densities
higher than a spin unpolarized one is a result of the exchange correlation
effects \cite{Dolgopolov}. Noting these results and supposing a similarity between ordinary matter and dark matter properties, one can assume that the spin-dependent interactions can also lead to a spontaneous spin polarization in a system of dark matter particles.
Therefore, it is
possible to take a system of DM particles into account which are spin polarized.
The spin polarization of DM particles could be a result of
the spin-dependent interactions and couplings to the nucleon \cite{Savvidy,Barnich,Tovey,Martins,Lin,Lopes11},
the spin-dependent interactions with nucleus \cite{Archambault,Martel,Martinez,Moulin}, the spin-spin interaction
 between DM particles \cite{Heo}, the dipole
moments of the WIMP with the conventional magnetism and also dark magnetism \cite{Chang},
the unconventional properties of DM which create
the galactic magnetic fields \cite{Berezhiani}, and the magnetic
field of quark-nuggets \cite{VanDevender}. As we explain in the following,
we focus on the influence of the spin polarized DM on the structure
of neutron stars. In this regard, one can anticipate that with a system of DM which
are spin polarized, the particles occupy the Fermi sphere in a way that the
Fermi momentum of the fermions is larger than the case of spin unpolarized one.
This effect can result in the more stiffening of the DM equation of state (EOS) and larger masses of neutron star.

Regarding the above discussions, the effects of DM particle spin on the physical properties
of star
can be considerable. Hereof, the importance of spin-dependent interaction in interferences \cite{Martinez}, the spin-dependent interactions as the dominant source of scattering \cite{Martinez}, the large spin induced cross section due to the Z-exchange \cite{Vergados},
and the decrease of central temperature in the Sun due to
DM with large spin-dependent interactions \cite{Cumberbatch} are some cases.
In this work, we are interested
in the properties of neutron stars which are affected by the spin polarized self-interacting DM.
Considering a system of DM particles with spin one half which can be spin polarized in the neutron star,
it is possible to understand the spin nature of DM particles, the bulk properties related to
the spin polarization of particles, and the strength of the interaction via the influence on the observational results.
Here we show how the spin and strength of interaction between DM particles affect the properties of neutron stars.

\section{Spin polarized self-interacting dark matter equation of state}
\label{sect:model}

We treat the DM in neutron star as an interacting spin polarized Fermi gas
at zero temperature. A system composed of $N$ particles with mass $m$ and spin $1/2$ is considered. The internal
energy per particle for this system is given by,
\begin{eqnarray}\label{etot}
           E_{tot}=E_1+E_2,
 \end{eqnarray}
in which $E_1$ and $E_2$ denote the one-body and interaction two-body energies. For the
spin polarized system, the one-body term is as follows,
\begin{eqnarray}
           E_{1}=\frac{1}{N}\sum_{i=+,-}\sum_{k\leqslant k_F^{(i)}}\sqrt{\hbar^2c^2k^2+m^2c^4},
            \end{eqnarray}
where $k_F^{(i)}$ is the Fermi momentum of a DM particle with spin projection $i$. It is easy to show that the one-body term takes the form,
\begin{eqnarray}
           E_{1}=\frac{m^4 c^5}{2 \pi^2 \hbar^3}\frac{1}{\rho}\sum_{i=+,-}\frac{1}{8}\{x_F^{(i)}\sqrt{1+{x_F^{(i)}}^2}(1+2{x_F^{(i)}}^2)-sinh^{-1}(x_F^{(i)})\}.
            \end{eqnarray}
In the above equation, $\rho$ is the
total number density of spin polarized DM particles and $x_F^{(i)}=\frac{\hbar k_F^{(i)}}{m c}$.
 \begin{figure*}[t]
\centering{%
{
  \label{speos}\includegraphics[width=1.0\textwidth]{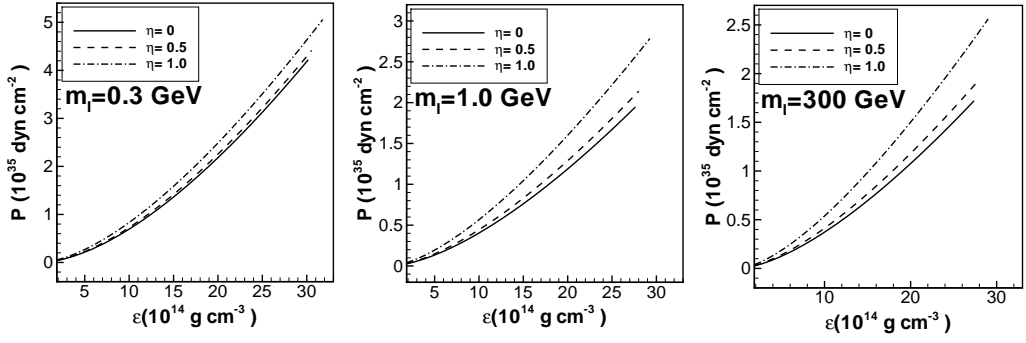}\qquad}}
\caption{Equation of state for the spin polarized DM with the mass $m=1 \ GeV$ and different values of interaction between particles, $m_I$, and spin polarization parameter, $\eta$.}
\label{fig1}
\end{figure*}
The interaction two-body energy in Eq. (\ref{etot}) can be expressed in the form \cite{Narain},
\begin{eqnarray}\label{e2pm}
           E_{2}=\frac{u}{\rho},
            \end{eqnarray}
in which $u$ is the interaction energy density of the particles. It should be noted that we have considered the spin-independent interaction between DM particles. In the lowest order
approximation, the interaction energy density is presented by \cite{Narain},
\begin{eqnarray}
          u=\frac{{\rho}^2}{m_I^2}.
            \end{eqnarray}
In the last equation, the value of $m_I$ shows the energy scale of the interaction between DM
particles \cite{Narain}.
Using this approximation, Eq. (\ref{e2pm}) leads to
\begin{eqnarray}
           E_{2}=\frac{\rho}{ m_I^2}.
            \end{eqnarray}
To quantify the amount of bulk spin polarization of DM, we introduce the parameter $\eta$ as follows,
\begin{eqnarray}
           \eta=\frac{\rho^{(+)}-\rho^{(-)}}{\rho},
            \end{eqnarray}
in which $\rho^{(i)}$ is the number density of DM particles with spin projection $i$.
Using this definition, the internal energy, Eq. (\ref{etot}), takes the form
\begin{eqnarray}
            E_{tot}&=&\frac{m}{16 \pi^2\rho}(m^2(3\pi^2\rho(1+\eta))^{1/3}\sqrt{1+\frac{(3\pi^2\rho(1+\eta))^{2/3}}{m^2}}
            \nonumber \\ &+& m^2(3\pi^2\rho(1-\eta))^{1/3}\sqrt{1+\frac{(3\pi^2\rho(1-\eta))^{2/3}}{m^2}}
              \nonumber \\&+&6\pi^2\rho(1+\eta)\sqrt{1+\frac{(3\pi^2\rho(1+\eta))^{2/3}}{m^2}}
             \nonumber \\&+&6\pi^2\rho(1-\eta)\sqrt{1+\frac{(3\pi^2\rho(1-\eta))^{2/3}}{m^2}}
            - m^3 sinh^{-1}(\frac{(3\pi^2\rho(1+\eta))^{1/3}}{m}) \nonumber \\&-&
            m^3 sinh^{-1}(\frac{(3\pi^2\rho(1-\eta))^{1/3}}{m}))
           +\frac{\rho}{ m_I^2}.
 \end{eqnarray}

In the next step, applying the first law of thermodynamics, $P=\rho^2 (\frac{\partial E_{tot}}{\partial \rho})$ , the pressure
of spin polarized DM is obtained.
Fig. \ref{fig1} presents the DM EOS related to the interacting spin polarized DM for the cases with different strength of interaction, $m_I$, and spin polarization parameter, $\eta$.
We can see that for each strength of the interaction, the stiffness of the EOS increases with the increase in the spin polarization parameter.
This increase in the stiffening of the DM EOS is due to the fact that with a system of more spin polarized DM, the particles occupy the Fermi sphere so that the
Fermi momentum of the fermions is larger.
The effects of spin polarizability of DM are more significant at higher densities. This Figure also confirms that the increase in the value of $m_I$,
which is corresponding to the decrease in the interaction between DM particles, leads to the softening of the EOS.

\section{Piecewise polytropic equation of state for the neutron star matter}
\begin{figure*}[t]
\centering{%
{
  \label{1}\includegraphics[width=0.5\textwidth]{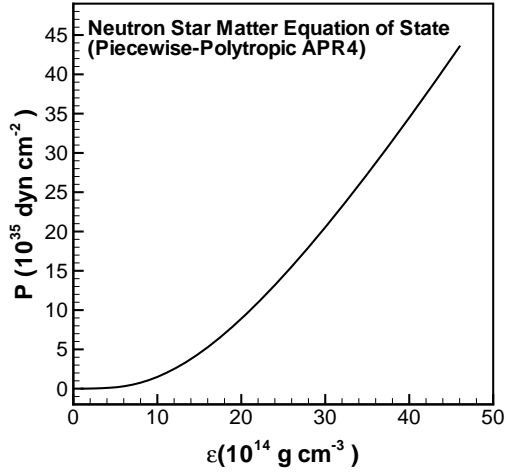}\qquad}}
\caption{Piecewise-polytropic EOS in APR4 model \cite{Akmal,Hotokezaka13}.}
\label{1}
\end{figure*}

To describe the neutron star matter (NSM) in our calculations, we apply a parameterized
piecewise-polytropic EOS \cite{Read,Hotokezaka}. For four segments of polytropes, the EOS for the NSM with the rest-mass density $b_i\leq b\leq b_{i+1}$, ($0\leq i\leq3$), is as follows,
\begin{eqnarray}
     P=K_i b^{\Gamma_i},
       \end{eqnarray}
in which P denotes the pressure,
$K_i$ shows the polytropic constant, and $\Gamma_i$ presents the adiabatic index.
In this model, the pressure is continuous at the boundaries of the piecewise polytropes, $b_i$. Similar to the parameters in \cite{Read}, we set $\Gamma_0= 1.3562395$ and $K_0=3.594\times10^{13}$ (cgs units). The values of the boundary density are also chosen
as $b_2=10^{14.7}\ g/cm^3$ and $b_3=10^{15.0}\ g/cm^3$. Therefore, with the free parameters ($P_2,\Gamma_1,\Gamma_2,\Gamma_3$) and the continuity of the pressure, the piecewise-polytropic EOS is completely determined. It should be noted that $P_2$ is the NSM pressure at $b=b_2$. In our calculations, we apply the piecewise-polytropic EOS in APR4 model with $log(P_2(dyn/cm^2))=34.269$, $\Gamma_1=2.830$, $\Gamma_2=3.445$, and $\Gamma_3=3.348$ \cite{Akmal,Hotokezaka13}. Moreover, using the first law of thermodynamics we have the following expression for the NSM energy density with the rest-mass density $b_i\leq b\leq b_{i+1}$ \cite{Read},
\begin{eqnarray}
     \epsilon(b)=(1+a_i)b+\frac{K_i}{\Gamma_i-1}b^{\Gamma_i},
       \end{eqnarray}
with
\begin{eqnarray}
     a_i=\frac{\epsilon(b_i)}{b_i}-1-\frac{K_i}{\Gamma_i-1}b_i^{\Gamma_i}.
       \end{eqnarray}
Fig. \ref{1} presents the piecewise-polytropic NSM EOS in APR4 model.
For the NSM with the number densities higher than $0.05\ fm^{-3}$, we apply the above EOS.
Besides, for the number densities lower than $0.05\ fm^{-3}$, the EOS calculated by Baym \cite{Baym} is used.

\section{Structure of neutron star with spin polarized self-interacting dark matter}

In order to study the structure of neutron stars which are admixed by the DM, we apply the two-fluid formalism \cite{Sandin,Ciarcelluti} for the NSM and DM.
In this model, a system is composed of NSM and DM particles which interact with each other just through gravity. We investigate the properties of a neutron star with two concentric
spheres. One of the spheres contains NSM, and the other
is formed by the spin polarized self-interacting DM. Considering the static and spherically symmetric space-time with the line element, (together with the units in which G = c = 1),
$d\tau^2=e^{2\nu(r)}dt^2-e^{2\lambda(r)}dr^2-r^2(d\theta^2+sin^2\theta d\phi^2)$,
and the energy-momentum tensor of a perfect fluid,
$T^{\mu \nu}=-p g^{\mu \nu}+(p+\varepsilon)u^{\mu}u^{\nu}$,
we obtain the structure of the neutron star. We denote the total
pressure and total energy density by $p$ and $\varepsilon$, respectively, which are related to the pressure and energy density of NSM and DM by
$p(r) = p_N(r) + p_{D}(r)$ and $ \varepsilon(r) =\varepsilon_N(r) + \varepsilon_{D}(r)$.
In this model,
the Einstein field equations lead to \cite{Sandin,Ciarcelluti,Xiang},
 \begin{eqnarray}
      e^{-2\lambda(r)} &=& 1-\frac{2M(r)}{r},\nonumber \\
        \frac{d\nu}{dr} &= &\frac{M(r)+4\pi r^3 p(r)}{r[r-2M(r)]},\nonumber \\
         \frac{dp_N}{dr} &=& -[p_N(r)+\varepsilon_N(r)] \frac{d\nu}{dr},\nonumber \\
          \frac{dp_{D}}{dr} &=& -[p_{D}(r)+\varepsilon_{D}(r)] \frac{d\nu}{dr},\nonumber \\
       \end{eqnarray}
in which $M(r)=\int_0^r dr 4 \pi r^2 \varepsilon(r)$ shows the total mass inside a sphere with radius $r$. The above relations are the result
of the assumption that two fluids interact just via gravity. These two-fluid TOV equations can be applied to calculate the neutron star structure and the properties of the NSM and DM spheres.
The radius, $R_N$, and mass, $M_N$, of NSM sphere are obtained with the condition $p_N(R_N) = 0$. In addition, the radius, $R_D$, and mass, $M_D$, of the DM sphere are given by the condition $p_{D}(R_{D})= 0$. The total mass of the neutron star is determined by the sum of the masses of NSM and DM spheres, i.e. $M=M_N+M_D$ . It is important to note that the
pressure and density profiles of two fluids are different with each other.

\subsection{Total mass and visible radius}

\begin{figure*}[t]
\centering{%
{
\label{Mtrn}   \includegraphics[width=1.0\textwidth]{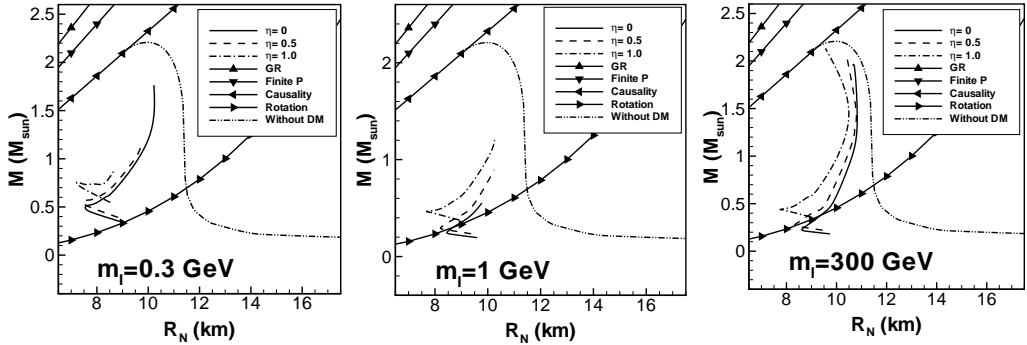}\qquad}}
\caption{Total mass of neutron star, $M$, with DM and without DM (normal neutron star) versus the radius of NSM sphere, $R_N$.
In addition, the curves which show the permitted region are presented, see the text.}
\label{fig3}
\end{figure*}

Fig. \ref{fig3} presents the total mass of neutron star with DM and without DM (Normal neutron star) versus
the visible radius, i.e. the radius of the NSM sphere. We have also shown the permitted
region by presenting the curves from the general relativity $M>\frac{c^2 R}{2 G}$ (GR),
finite pressure $M>\frac{4}{9}\frac{c^2 R}{G}$ (Finite P), causality $M>\frac{10}{29}\frac{c^2 R}{G}$ (Causality),
and rotation of 716 Hz pulsar J1748-2446ad (Rotation) \cite{Lattimer}. It is clear that the existence of DM in the neutron star leads to the reduction of neutron star size. In addition, the maximum mass of neutron star with DM is lower than this quantity for the normal neutron star. In the most cases, for each interaction strength, the visible size of neutron star with a specific mass
reduces by increasing the polarization of DM. Therefore, the more polarized
the DM, the more compact the neutron star.
In addition, we can see that the neutron
stars with full polarized DM could have lower size. This is while by
decreasing the polarization of DM, the smaller neutron stars are not acceptable.
For stars with a special size, the mass of neutron star grows when the spin polarization parameter
increases. This is a consequence of the higher Fermi momentum of the fermions and more stiff DM EOSs in the cases with the more spin polarized DM.
 Comparing the neutron stars with DM and without DM indicates that with DM, regardless of its polarization and strength of the interaction, the behavior of the total mass versus the visible radius, i.e. $M-R_N$ relation, is essentially different from this relation for the normal neutron
stars.
\subsection{Total mass versus the radius of dark matter sphere}

\begin{figure*}[t]
\centering{%
{ \label{Mtrd}  \includegraphics[width=1.0\textwidth]{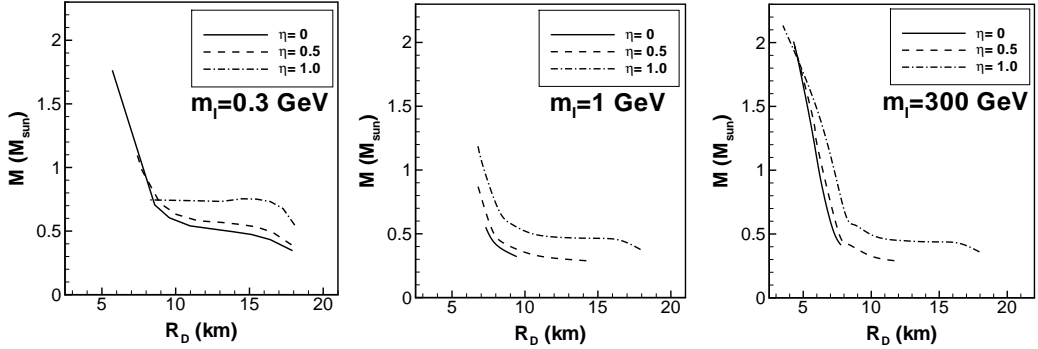}\qquad}}
\caption{Total mass of neutron star, $M$, versus the radius of DM sphere, $R_{D}$.}
\label{fig5}
\end{figure*}

Fig. \ref{fig5} presents the total mass of neutron star
versus the radius of DM sphere. In all cases, the DM sphere is smaller for more massive
stars.
We can see that for $m_I=1\ GeV$ and $m_I=300\ GeV$, the increase in the spin
polarization parameter of DM leads to the larger size of
the DM sphere.
The increase in the size of DM sphere due to the polarization of DM
is more significant in the case with $\eta=1$. The increase in the mass of the stars with the same size DM sphere is clear when the spin polarization parameter grows; another result of the more stiff DM EOS.
Figs. \ref{fig3} and \ref{fig5} confirm that for more massive stars, the radius of NSM sphere is larger than
the radius of DM sphere.
This shows that the visible matter surrounds
the DM sphere.
However, in the stars with lower masses, the DM sphere
has a size bigger than the NSM sphere.
In the cases of massive stars, the radii of NSM sphere and DM sphere
can be $11\ km$ and $4\ km$, respectively.
However, for the low mass stars, these radii
can reach $7\ km$ and $18\ km$. Thus, the DM sphere can vary in size more than the NSM sphere.

\subsection{Neutron star matter sphere mass versus the visible radius}

\begin{figure*}[t]
\centering{%
{
  \label{Mnrn} \includegraphics[width=1.0\textwidth]{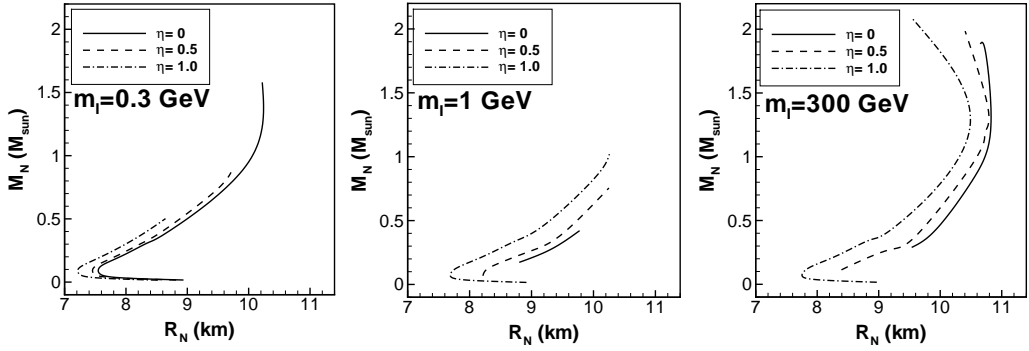}\qquad}}
\caption{Mass of NSM sphere, $M_N$, as a function of NSM sphere radius, $R_N$.}
\label{fig9}
\end{figure*}

Fig. \ref{fig9} presents the contribution of NSM portion in the mass of star, $M_N$,
versus the radius of NSM sphere.
For the stars with
larger NSM sphere, the mass of NSM sphere is higher. It is clear that
with a mass for NSM sphere, the NSM sphere radius decreases by increasing
the spin polarization parameter. Fig. \ref{fig9} also confirms that for the stars with the same visible size, the mass of NSM sphere grows by increasing the spin polarization parameter of DM. It means that the stiffer DM EOS shifts the mass of NSM sphere to higher values.
Figs. \ref{fig3} and \ref{fig9} verify that the behavior of the NSM sphere mass and total mass
versus the size of NSM sphere are similar.
\subsection{Mass and radius of dark matter sphere}
\begin{figure*}[t]
\centering{%
{
 \label{Mdrd}  \includegraphics[width=1.0\textwidth]{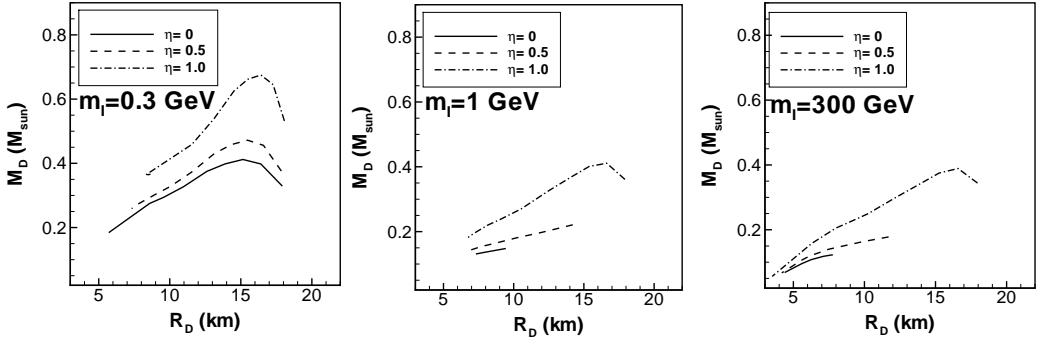}\qquad}}
\caption{Mass of DM sphere, $M_{D}$, versus its radius, $R_{D}$.}
\label{fig11}
\end{figure*}

Fig. \ref{fig11} shows the mass of DM sphere versus its radius.
For the stars with larger DM sphere, the contribution of this sector
in the total mass is more considerable.
The range of mass of this sphere
is $0.06M_{\odot} \lesssim M_{D} \lesssim 0.68M_{\odot}$ which is smaller than the range $0.01M_{\odot} \lesssim M_N \lesssim 2.08M_{\odot}$ related to the NSM sphere (see Fig. \ref{fig9}).
Fig. \ref{fig11} indicates
that by increasing the value of $m_I$, the $M_{D}-R_{D}$ relation becomes more flattened
and the mass contribution of DM decreases.
This is while for each value of the interaction strength, the contribution of DM
in the total mass grows by increasing the spin polarization parameter; another consequence of stiffer DM EOSs.
This indicates that the star with more spin polarized DM have more massive
halo of DM.
The radius of DM sphere lies between $4\ km \lesssim R_{D} \lesssim 18\ km$ which is a wider range
in comparison with $7\ km\lesssim R_N \lesssim 11\ km$ for the NSM sphere.
This phenomenon, along with the one related to the
mass of spheres results in the less accumulation of DM in comparison
with the visible matter.

\subsection{Gravitational redshift versus the total mass}
\begin{figure*}[t]
\centering{%
{
 \label{z}  \includegraphics[width=0.9\textwidth]{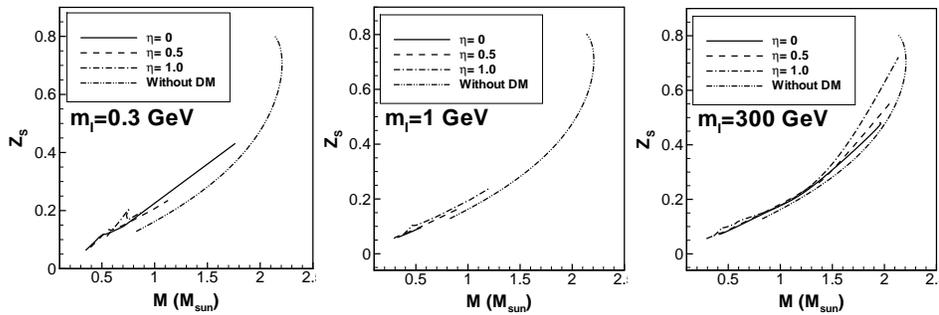}\qquad}}
\caption{Gravitational redshift at the surface of neutron star, $Z_S$, with DM and without DM (normal neutron star).}
\label{fig21}
\end{figure*}

We have shown the gravitational redshift at the surface of neutron star with DM and without DM (normal neutron star) in Fig. \ref{fig21}. For a special mass, the gravitational redshift of neutron star with DM is higher compared to the normal neutron star.
In most cases, with a certain mass of neutron star, the gravitational redshift increases with the increase
in the spin polarization parameter. In addition, the strength of interaction between DM particles affects the gravitational redshift.
Fig. \ref{fig21} indicates that the influence of spin polarizability on the gravitational redshift is more significant for the case with $m_I=300\ GeV$, i.e. with lower self-interaction of DM particles.
Since the gravitational redshift is obtained using the observational data,
it is possible to use our results to estimate the interaction and also polarization of DM.

\section{Conclusion}

Employing the piecewise polytropic equation of state for the neutron star matter and the equation of state of spin polarized self-interacting dark matter, the properties of neutron stars affected by the spin polarized self-interacting
dark matter have been considered. The general relativistic formalism and
the spin polarized self-interacting dark matter equation of state have been employed.
The increase in the spin polarization parameter of the dark matter results
in the more stiffening of the EOS.
Our results show that the visible size of the neutron star decreases
by increasing the spin polarization of DM.
This is while, the size of the dark matter sphere increases with the growth of spin polarization parameter.
Because of the stiffer DM EOSs in the cases with more spin polarized DM,
the total mass of neutron star, the mass of NSM sphere, and the mass of DM sphere, grow when the spin polarization parameter increases.
It has been confirmed that the relations $M-R_N$, $M-R_D$, $M_N-R_N$, and $M_D-R_D$ are influenced by the spin polarization parameter of DM.
We have shown that the behavior of the NSM sphere mass and total mass
versus the size of NSM sphere are similar.
Moreover, the gravitational redshift of neutron star grows when the spin polarization parameter
increases.
\section{Acknowledgments}
The author wishes to thank the Shiraz University Research
Council.

\end{document}